\documentclass[a4paper,10pt]{article}
\usepackage[dvips]{graphicx}
\usepackage{amssymb,amsmath}
\numberwithin{equation}{section}

\begin{document}

 \def\gsim{ \lower .75ex \hbox{$\sim$} \llap{\raise .27ex \hbox{$>$}} }
 \def\lsim{ \lower .75ex \hbox{$\sim$} \llap{\raise .27ex \hbox{$<$}} }



 \title{\Large DARBOUX TRANSFORMATION AND SOLUTIONS OF THE (2+1)-DIMENSIONAL SCHR\"ODINGER-MAXWELL-BLOCH EQUATION}

\author{ G. Shaikhova, K. Yesmakhanova, G. Mamyrbekova and  R.Myrzakulov\footnote{The corresponding author.
Email: rmyrzakulov@gmail.com}
 \\ \textit{Eurasian International Center for Theoretical Physics and  Department of General } \\ \textit{ $\&$  Theoretical Physics, Eurasian National University, Astana 010008, Kazakhstan}}

\date{}
 \maketitle


 \renewcommand{\baselinestretch}{1.1}

 \begin{abstract}
In this paper, we construct a Darboux transformation (DT) of the (2+1)-dimensional Schr\"odinger-Maxwell-Bloch equation (SMBE) which is integrable by the Inverse Scattering Method. Using this DT, the one-soliton solution and periodic solution are obtained from the "seed" solutions. 
 \end{abstract}

 
\section{Introduction}
It is well known that  the nonlinear nature of the real system
is considered to be fundamental in modern science. Nonlinearity is the  fascinating subject which 
has many applications  in almost all areas of science. Usually  nonlinear phenomena are    modeled by nonlinear odinary and/or  partial differential  equations. Many of these nonlinear  differential  equations (NDE) are  completely integrable. This means that these integrable NDE have some class interesting exact solutions such as  solitons, dromions, rogue waves, similaritons and so on. They are of great mathematical
as well as physical interest and the investigation  of the solitons  and other its sisters  have become one of the most exciting and extremely
active areas of research in modern science and technology in the past several  decades.
In particular, many of the completely integrable
NDE  are found and studied.
Among of such integrable nonlinear systems the Schr\"odinger-Maxwell-Bloch  equation (SMBE) plays an important role. The SMBE describe a soliton propagation in fibres with resonant and erbium-doped systems \cite{Mai} and has the (1+1)-dimensions. In this paper our aim is to construct the Darboux tranformation (DT) for the (2+1)-dimensional SMBE and finding its soliton solutions.

The paper is organized as follows.  In Sec. 2,  some main notations of our models  are introduced. As an example, we consider the (1+1)-dimensional SMBE. The  (2+1)-dimensional SMBE we present in Sec. 3. The DT of the  (2+1)-dimensional SMBE we construct in Sec. 4. In the next Sec. 5, some exact solutions of the  (2+1)-dimensional SMBE are given. Section 6 is devoted to conclusions.

\section{The Schr\"odinger-Maxwell-Bloch  equation in 1+1 dimensions}

In this paper, we will study some properties of the (2+1)-dimensional SMBE. But in order to be self-contained, in this section we shall recall some important informations on the (1+1)-dimensional SMBE.  In 1+1 dimensions, the SMBE is given by \cite{Porsezian1}
\begin{eqnarray}
iq_{t}+q_{xx}+2\delta |q|^2q-2ip&=&0, \label{2.1}\\
p_{x}-2i\omega p -2\eta q&=&0,\label{2.2}\\
\eta_{x}+\delta(q^{*} p +p^{*} q)&=&0,\label{2.3}
\end{eqnarray}
where $q, p$ are complex functions, $\eta$ is a real function, $\omega, \delta$ are real contants ($\delta=\pm 1$) \cite{R13}. If  $\delta=+1$ $(\delta=-1)$ then we get the SMBE with the attractive (repulsive) interaction. This   (1+1)-dimensional SMBE is integrable  (see e.g. \cite{C1}-\cite{C2} and references therein). Its Lax pair has the form
\begin{eqnarray}
\Psi_{x}&=&U\Psi,\label{2.4}\\
\Psi_{t}&=&2\lambda U\Psi+B\Psi,\label{2.5} 
\end{eqnarray}  
where 
\begin{eqnarray}
U&=&-i\lambda \sigma_3+U_0,\label{2.6}\\
B&=&B_0+\frac{i}{\lambda+\omega}B_{-1}.\label{2.7}
\end{eqnarray}
Here
\begin{eqnarray}
U_0&=&\begin{pmatrix} 0&q\\-\delta q^{*}& 0\end{pmatrix},\label{2.8}\\
B_0&=&i\delta|q|^2\sigma_3+i\begin{pmatrix} 0&q_x\\\delta q^{*}_x& 0\end{pmatrix},\label{2.9}\\
B_{-1}&=&\begin{pmatrix} \eta&-p\\-\delta p^{*}& -\eta\end{pmatrix}.\label{2.10} 
\end{eqnarray}
 As well-known,  the self-induced transparency
 effect  in erbium doped nonlinear fibre in two-level resonant
media describes by the SMBE [19,23]. In the literature, many types exact solutions of the SMB equation (\ref{2.1})-(\ref{2.3}) have been found (see e.i. \cite{He1}).  

Another interesting integrable system related with the SMBE (\ref{2.1})-(\ref{2.3}) is the so-called  M-XCIX equation. The M-XCIX  equation reads as  \cite{R13}
\begin{eqnarray}
{\bf S}_{t}+{\bf S}\wedge {\bf S}_{xx}+\frac{2}{\omega}{\bf S}\wedge {\bf W}&=&0,\label{2.11}\\
 {\bf W}_{x}+2\omega {\bf S}\wedge  {\bf W}&=&0,\label{2.12} \end{eqnarray} 
 where $\wedge$ denotes a vector product and \begin{equation}
{\bf S}=(S_1,S_2,S_3), \quad {\bf W}=(W_1,W_2,W_3).\label{2.13} \end{equation} 
  Here $\omega$ is a real function, ${\bf S}^2=S_{1}^2+S_{2}^2+S_{3}^2=1$, $S_i$ and $W_i$ are some real functions. The M-XCIX equation is integrable.  Its Lax representation has the form 
\begin{eqnarray}
\Phi_{x}&=&U\Phi,\label{2.14}\\
\Phi_{t}&=&V\Phi,\label{2.15} 
\end{eqnarray}  
where
 \begin{eqnarray}
U&=&-i\lambda S,\label{2.16}\\
V&=&\lambda^2V_2+\lambda V_{1}+\frac{i}{\lambda+\omega}V_{-1}-\frac{i}{\omega}V_{0}.\label{2.17} 
\end{eqnarray} 
Here
\begin{eqnarray}
V_2&=&-2i S,\label{2.18}\\
V_1&=&0.5[S,S_x],\label{2.19}\\
V_{-1}&=&V_0=\begin{pmatrix} W_3&W^{+}\\W^{-}& -W_3\end{pmatrix},\label{2.20} 
\end{eqnarray} 
where
 \begin{equation}\label{2.21}
S=S_i\sigma_i=\begin{pmatrix} S_3 &S^{-}\\S^{+} & -S_3\end{pmatrix}, \quad W=W_i\sigma_i=\begin{pmatrix} W_3 &W^{+}\\W^{-} & -W_3\end{pmatrix}.
 \end{equation}
Here $S^{\pm}=S_1\pm iS_2, \quad W^{\pm}=W_1\pm i W_2,$  $[A,B]=AB-BA, $ $\sigma_i$ are Pauli matrices.
With such $U,V$ matrices,  the equation 
\begin{equation}\label{2.22}
U_t-V_x+[U,V]=0
 \end{equation}
is equivalent to the equation \begin{eqnarray}
iS_{t}+0.5[S, S_{xx}]+\frac{1}{\omega}[S, W]&=&0,\label{2.23}\\
 iW_{x}+\omega [S, W]&=&0,\label{2.24} \end{eqnarray}
 It is nothing but the matrix form of the  M-XCIX equation (\ref{2.11})-(\ref{2.12}). Finally we recall that if $p=\eta=W_i=0$ then the SMBE (\ref{2.1})-(\ref{2.3}) and  the M-XCIX equation (\ref{2.11})-(\ref{2.12}) reduce to the NLSE 
 \begin{eqnarray}
iq_{t}+q_{xx}+2\delta |q|^2q=0\label{2.25}
\end{eqnarray}
 and to the Heisenberg ferromagnet equation
 \begin{eqnarray}
{\bf S}_{t}+{\bf S}\wedge {\bf S}_{xx}=0,\label{2.26} \end{eqnarray}
 respectively.

\section{The (2+1)-dimensional  Schr\"odinger-Maxwell-Bloch equation}
 Our aim in this paper is to construct the DT for the (2+1)-dimensional SMBE and using this to find some important its exact solutions. The (2+1)-dimensional SMBE reads as \cite{R14}
  \begin{eqnarray}
iq_{t}+q_{xy}-vq-2ip&=&0, \label{3.1}\\
ir_{t}-r_{xy}+vr-2ik&=&0,\label{3.2}\\
v_{x}+2(rq)_{y}&=&0,\label{3.3}\\
p_{x}-2i\omega p -2\eta q&=&0,\label{3.4}\\
k_x+2i\omega k-2\eta r&=&0,\label{3.5}\\
\eta_{x}+r p +k q&=&0,\label{3.6}
\end{eqnarray}
where $q, k, r, v,  p$ are complex functions, $\eta$ is a real function, $\omega$ is the  real contant. It is integrable by the IST. The corresponding  Lax representation is given by
\begin{eqnarray}
\Psi_{x}&=&A\Psi,\label{3.7}\\
\Psi_{t}&=&2\lambda\Psi_y+B\Psi,\label{3.8} 
\end{eqnarray}  
where $A$ and $B$ have the form 
 \begin{eqnarray}
A&=&-i\lambda \sigma_3+A_0,\label{3.9}\\
B&=&B_0+\frac{i}{\lambda+\omega}B_{-1}.\label{3.10} 
\end{eqnarray} 
Here
\begin{eqnarray}
A_0&=&\begin{pmatrix} 0&q\\-r& 0\end{pmatrix},\label{3.11}\\
B_0&=&-0.5iv\sigma_3+i\begin{pmatrix} 0&q_y\\r_y& 0\end{pmatrix},\label{3.12}\\
B_{-1}&=&\begin{pmatrix} \eta&-p\\-k& -\eta\end{pmatrix}.\label{3.13} 
\end{eqnarray}
Let us now  we consider the reduction $r=\delta q^{*}, \quad k=\delta p^{*}$, where $*$ means a complex conjicate and $\delta$ is the  real contant. Then the system (\ref{3.1})-(\ref{3.6}) takes the form
 \begin{eqnarray}
iq_{t}+q_{xy}-vq-2ip&=&0, \label{3.14}\\
v_{x}+2\delta(|q|^2)_{y}&=&0,\label{3.15}\\
p_{x}-2i\omega p -2\eta q&=&0,\label{3.16}\\
\eta_{x}+\delta(q^{*} p +p^{*} q)&=&0. \label{3.17}
\end{eqnarray}
Here  ($\delta=\pm 1$) so that  $\delta=+1$ corresponds to the attractive interaction and  $\delta=- 1$ to the repulsive interaction respectively.  Let us we also present the spin system which is the equivalent counterpart of the (2+1)-dimensional SMBE (\ref{3.14})-(\ref{3.17}). It looks like \cite{R14}
\begin{eqnarray}
iS_{t}+\frac{1}{2}([S, S_{y}]+2iuS)_{x}+\frac{1}{\omega}[S, W]&=&0,\label{3.18}\\
u_xI-\frac{i}{2}S[S_x,S_y]&=&0,\label{3.19}\\
 iW_{x}+\omega [S, W]&=&0\label{3.20}
\end{eqnarray} 
or equivalently
\begin{eqnarray}
iS_{t}+\frac{1}{2}[S, S_{xy}]+iuS_{x}+\frac{1}{\omega}[S, W]&=&0,\label{3.21}\\
u_x-\frac{i}{4}tr(S[S_x,S_y])&=&0,\label{3.22}\\
 iW_{x}+\omega [S, W]&=&0.\label{3.23}
\end{eqnarray} 
It is  the so-called   ML-II equation \cite{R14}. 
 The ML-II equation (\ref{3.18})-(\ref{3.20}) is integrable by the IST.  Its   Lax representation can be written in the form
 \begin{eqnarray}
\Phi_{x}&=&U\Phi,\label{3.24}\\
\Phi_{t}&=&2\lambda\Phi_y+V\Phi.\label{3.25} 
\end{eqnarray}  
Here the matrix operators $U$ and $V$ have  the form 
 \begin{eqnarray}
U&=&-i\lambda S,\label{3.26}\\
V&=&\lambda V_{1}+\frac{i}{\lambda+\omega}W-\frac{i}{\omega}W,\label{3.27} 
\end{eqnarray} 
where
\begin{eqnarray}
V_1&=&2Z=\frac{1}{2}([S, S_{y}]+2iuS),\label{3.28}\\
W&=&\begin{pmatrix} W_3&W^{-}\\W^{+}& -W_3\end{pmatrix}.\label{3.29} 
\end{eqnarray}  
Some comments in order. First we recall  that for these SMBE and ML-II equation,  the spectral parameter obeys the equation: $\lambda_t=2\lambda\lambda_y$ which has for example the following particular solution: $\lambda=(\beta_1+\beta_2y)(\beta_3-2\beta_1t)^{-1}$, where $\beta_j$ are  in general some complex constants.
 Secondly, we also note that in 1+1 dimensions that is if $y=x$, both equations take the form  of the (1+1)-dimensional SMBE (\ref{2.1})-(\ref{2.3}) and the M-XCIX equation (\ref{2.11})-(\ref{2.12}) respectively. 
 Third, if $p=\eta=W_i=0$ then the (2+1)-dimensional SMBE (\ref{2.1})-(\ref{2.3}) and the ML-II  equation (\ref{2.11})-(\ref{2.12}) reduce to the (2+1)-dimensional nonlinear Schr\"odinger equation of the form
 \begin{eqnarray}
iq_{t}+q_{xy}-vq&=&0, \label{3.30}\\
v_{x}+2\delta(|q|^2)_{y}&=&0\label{3.31}
\end{eqnarray}
 and to the M-I equation \cite{Chen}
 \begin{eqnarray}
iS_{t}+\frac{1}{2}[S, S_{xy}]+iuS_{x}&=&0,\label{3.32}\\
u_x-\frac{i}{4}tr(S[S_x,S_y])&=&0,\label{3.33}
\end{eqnarray}
 respectively. Some properties of some integrable and notintegrable spin systems were studied in the refs. \cite{royal}-\cite{myrzakulov-1397}.

 \section{Darboux transformation}

It is well-known that the DT has been proved to be an efficient way to find the exact solutions like solitons, dromions, positons, breathers, rogue wave solutions  for integrable equations in 1+1 and 2+1 dimensions. In this section, considering the particularity of the Lax representation, we construct the DT of the (2+1)-dimensional SMBE (\ref{3.14})-(\ref{3.17}). Furthermore, we will find some  solutions of the (2+1)-dimensional SMBE using its DT. 

\subsection{One-fold DT}
We consider the following transformation of Eq.(\ref{3.7})-(\ref{3.8})
\begin{eqnarray}
\Psi'=T\Psi=(\lambda I-M)\Psi \label{4.1}
\end{eqnarray}
such that
\begin{eqnarray}
\Psi'_{x} &=& A'\Psi',\label{4.2}\\
\Psi'_{t} &=& 2\lambda\Psi'_{y}+B'\Psi', \label{4.3}
\end{eqnarray}
where $A'$ and $B'$ depend on $q'$, $v'$, $p'$, $\eta'$ and $\lambda$. Here 
\begin{eqnarray}
M=\begin{pmatrix} m_{11}&m_{12}\\m_{21}&m_{22}\end{pmatrix},\quad I=\begin{pmatrix} 1&0\\ 0&1\end{pmatrix}. \label{4.4}
\end{eqnarray}
The  relation between   $q'$, $v'$, $p'$, $\eta'$,  $\lambda$ and  $A'$ - $B'$ is the same as the relation between  $q$, $v$, $p$, $\eta$, $\lambda$ and  $A$-$B$. In order to hold Eqs.(\ref{4.2})-(\ref{4.3}), the $T$ must satisfies the following equations
\begin{eqnarray}
T_{x}+TA &=& A'T,\label{4.5} \\
T_{t}+TB &=& 2\lambda T_{y}+B'T. \label{4.6}
\end{eqnarray}
Then the relation between $q$, $v$, $p$, $\eta$  and $q'$, $v'$, $p'$, $\eta'$ can be reduced from these equations, which is in fact the DT of the (2+1)-dimensional SMBE.
Comparing the coefficients of $\lambda^{i}$ of the two sides of the equation (\ref{4.5}), we get
   \begin{eqnarray}
		\lambda^0&:&  M_{x}=A'_{0}M-MA_0, \label{4.7}\\
	 \lambda^1&:& A'_{0}=A_{0}+ i[M,\sigma_{3}], \label{4.8}\\
 \lambda^2&:& iI\sigma_{3}=i\sigma_{3}I. \label{4.9}
\end{eqnarray}
From (\ref{4.8}) we obtain 
\begin{eqnarray}
q^{[1]} &=& q-2im_{12}, \label{4.10}\\
r^{[1]} &=& r-2im_{21}. \label{4.11}
\end{eqnarray}
or\begin{eqnarray}
q^{[1]} &=& q-2im_{12}, \label{4.12}\\
\delta {q^{*}}^{[1]} &=& \delta q^{*}-2im_{21}.  \label{4.13}
\end{eqnarray}
Hence we get $m_{21}=- m_{12}^{*}$ in our    attractive interaction case that is if $\delta=+1$. Then comparing the coefficients of $\lambda^{i}$ of the two sides of the  equation (\ref{4.6}) gives us 
\begin{eqnarray} 
\lambda^{0}&:& -M_{t}=iB'_{0}-B'_{0}M-iB_{-1}+MB_{0}, \label{4.14}\\
\lambda^{1}&:& 2M_{y}=B'_{0}-B_{0}, \label{4.15}\\
(\lambda+\omega)^{-1}&:&0=-i\omega B'_{-1}-iB'_{-1}M+i\omega B_{-1}+iMB_{-1}. \label{4.16}
\end{eqnarray}
The last equation of this system gives
\begin{eqnarray}
B'_{-1}=(M+\omega I)B_{-1}(M+\omega I)^{-1}. \label{4.17}
\end{eqnarray}
At the same, from Eq.(\ref{4.15}) we get 
\begin{eqnarray}
v'=v+4im_{11y}=v-4im_{22y}. \label{4.18}
\end{eqnarray}
and hence we additionally  have  $m_{22}=m_{11}^{*}$. So the matix $M$ has the form
\begin{equation}
M=\begin{pmatrix} m_{11} & m_{12}\\ -m_{12}^{*} & m_{11}^{*}\end{pmatrix}, \quad M^{-1}=\frac{1}{|m_{11}|^2+|m_{12}|^{2}}\begin{pmatrix} m_{11}^{*} & -m_{12}\\ m_{12}^{*} & m_{11}\end{pmatrix},  \label{4.19}
\end{equation}
\begin{equation}
M+\omega I=\begin{pmatrix} m_{11}+\omega & m_{12}\\ -m_{12}^{*} & \omega+m_{11}^{*}\end{pmatrix},\quad 
(M+\omega I)^{-1}=\frac{1}{\square}\begin{pmatrix} m_{11}^{*}+\omega & -m_{12}\\ m_{12}^{*} & \omega+m_{11}\end{pmatrix}. \label{4.20}
\end{equation}
Here
\begin{equation}
\square=det(M+\omega I)=\omega^2+\omega(m_{11}+m_{11}^{*})+|m_{11}|^{2}+|m_{12}|^{2}. \label{4.21}
\end{equation}
The equation (\ref{4.17}) gives  
\begin{eqnarray}
\eta'&=& \frac{(|\omega+m_{11}|^{2}-|m_{12}|^{2})\eta-pm_{12}^{*}(\omega+m_{11})- p^{*}m_{12}(\omega+m_{11}^{*})}{\square}, \label{4.22}\\
p'&=& \frac{p(\omega+m_{11})^{2}-p^{*}m_{12}^{2}+2\eta m_{12}(\omega+m_{11})}{\square},\\ \label{4.23}
{p^{*}}'&=&\frac{p^{*}(\omega+m_{11}^{*})^{2}-pm_{12}^{*2}
+2\eta m_{12}^{*}(\omega+m_{11}^{*})}{\square}. \label{4.24}
\end{eqnarray}
 We now assume that 
\begin{eqnarray}
M=H\Lambda H^{-1}, \label{4.25}
\end{eqnarray}
where \begin{eqnarray}
H=\begin{pmatrix} \psi_{1}(\lambda_{1};t,x,y)&\psi_{1}(\lambda_{2};t,x,y)\\\psi_{2}(\lambda_{1};t,x,y)&\psi_{2}(\lambda_{2};t,x,y)\end{pmatrix}. \label{4.26}
\end{eqnarray}
Here
\begin{eqnarray}
\Lambda&=&\begin{pmatrix} \lambda_{1}&0\\0&\lambda_{2}\end{pmatrix} \label{4.27}
\end{eqnarray}
and $det$ $H\neq0$, where $\lambda_{1}$ and $\lambda_2$ are complex constants.
Using Eqs.(\ref{3.7})-(\ref{3.8}), it is easy to get
\begin{eqnarray}
H_{x} &=& -i\sigma_{3}H\Lambda+A_{0}H, \label{4.28}\\
H_{t} &=& 2 H_{y}\Lambda +B_{0}H+B_{-1}H\Sigma, \label{4.29}
\end{eqnarray}
where 
\begin{eqnarray}
\Sigma=\begin{pmatrix} \frac{i}{\lambda_{1}+\omega}&0\\0&\frac{i}{\lambda_{2}+\omega}\end{pmatrix}. \label{4.30}
\end{eqnarray}
  In order to satisfy the constraints of $S$ and  $B'_{-1}$ as mentioned above, we first notes that if $\delta=1$ then
\begin{eqnarray}
\Psi^{+}=\Psi^{-1}, \quad A_{0}^{+}=-A_{0}, \label{4.31}
\end{eqnarray} 
\begin{eqnarray}
\lambda_{2}=\lambda^{*}_{1}, \quad
 H=\begin{pmatrix} \psi_{1}(\lambda_{1};t,x,y)&-\psi^{*}_{2}(\lambda_{1};t,x,y)\\ \psi_{2}(\lambda_{1};t,x,y)&\psi^{*}_{1}(\lambda_{1};t,x,y)\\ \end{pmatrix}, \label{4.32}
\end{eqnarray}
\begin{eqnarray}
H^{-1}=\frac{1}{\Delta}\begin{pmatrix} \psi^{*}_{1}(\lambda_{1};t,x,y)&\psi^{*}_{2}(\lambda_{1};t,x,y)\\ -\psi_{2}(\lambda_{1};t,x,y)&\psi_{1}(\lambda_{1};t,x,y)\\ \end{pmatrix}. \label{4.33}
\end{eqnarray}
So for the matrix $M$ we have
\begin{eqnarray}
M&=&\frac{1}{\Delta}\begin{pmatrix} \lambda_{1}|\psi_{1}|^2+\lambda_{2}|\psi_{2}|^2 & (\lambda_{1}-\lambda_{2})\psi_{1}\psi_{2}^{*}\\ (\lambda_{1}-\lambda_{2})\psi_{1}^{*}\psi_{2} & \lambda_{1}|\psi_{2}|^2+\lambda_{2}|\psi_{1}|^2)\end{pmatrix},  \label{4.34}
\end{eqnarray}
where 
\begin{eqnarray}
\Delta &=&|\psi_{1}|^2+|\psi_{2}|^2. \label{4.35}
\end{eqnarray}
Here we mention  that $m_{22}=m_{11}^{*}$ and $m_{21}=-m_{12}^{*}$ that holds if $\lambda_2=\lambda_{1}^{*}$. So finally we get the following DT of the (2+1)-dimensional SMBE:
\begin{eqnarray}
q^{[1]}& = &q-2im_{12}=q-\frac{2i(\lambda_{1}-\lambda_{2})\psi_{1}\psi^{*}_{2}}{\Delta}, \label{4.36}\\
v^{[1]}& = &v+4im_{11y}=v-4i\left(\frac{\lambda_{1}|\psi_{1}|^2+\lambda_{2}|\psi_{2}|^2}{\Delta}\right)_{y}, \label{4.37}\\
\eta^{[1]}&=& \frac{(|\omega+m_{11}|^{2}-|m_{12}|^{2})\eta-pm_{12}^{*}(\omega+m_{11})- p^{*}m_{12}(\omega+m_{11}^{*})}{\square}, \label{4.38}\\
p^{[1]}&=& \frac{p(\omega+m_{11})^{2}-p^{*}m_{12}^{2}+2\eta m_{12}(\omega+m_{11})}{\square}, \label{4.39}\\
{p^{*}}^{[1]}&=&\frac{p^{*}(\omega+m_{11}^{*})^{2}-pm_{12}^{*2}
+2\eta m_{12}^{*}(\omega+m_{11}^{*})}{\square}. \label{4.40}
\end{eqnarray}
For our futher convenience, it is useful  rewrite these formulas in the determinant form. First we have 
\begin{eqnarray}
m_{11}=\frac{\lambda_{1}|\psi_{1}|^2+\lambda_{2}|\psi_{2}|^2}{\Delta}=\frac{\Delta_{11}}{\Delta}, \quad 
m_{12}=\frac{(\lambda_{1}-\lambda_{2})\psi_{1}\psi_{2}^{*}}{\Delta}=\frac{\Delta_{12}}{\Delta}, \label{4.41}
\end{eqnarray}
where
\begin{eqnarray}
\Delta_{11}=det\begin{pmatrix} \psi_{1} & -\lambda_{2}\psi_{2}^{*}\\ 
\psi_{2} & \lambda_{1}\psi_{1}^{*}\end{pmatrix}, \quad \Delta_{12}=-det\begin{pmatrix} \psi_{1} & \lambda_{1}\psi_{1}\\ 
\psi_{2}^{*} & \lambda_{2}\psi_{2}^{*}\end{pmatrix}. \label{4.42}
\end{eqnarray}
Finally we note that we can rewrite the above presented solution in terms of the elements of the matrix $T$. From 
\begin{eqnarray}
T=\lambda I -M=\lambda I +t^{[1]}_{0}=\frac{1}{\Delta_{1}}\begin{pmatrix} T_{111} & T_{112}\\ T_{121} & T_{122}\end{pmatrix} \label{4.43}
\end{eqnarray}
follows that
\begin{eqnarray}
M=\begin{pmatrix} m_{11} & m_{12}\\ -m_{12}^{*} & m_{11}^{*}\end{pmatrix}=\lambda I -T= -t^{[1]}_{0}.  \label{4.44}
\end{eqnarray}
Finally we can write the 1-sd solution of the (2+1)-dimensional SMBE which follows from the corresponding  one-fold DT, as
\begin{eqnarray}
q^{[1]}& = &q+2i(t^{1}_{0})_{12}, \label{4.46}\\
v^{[1]}& = &v-4i(t^{1}_{0})_{11y}, \label{4.47}\\
\eta^{[1]}&=& \frac{(|\omega-(t^{1}_{0})_{11}|^{2}-|(t^{1}_{0})_{12}|^{2})\eta+p(t^{1}_{0})_{12}^{*}(\omega-(t^{1}_{0})_{11})+ p^{*}(t^{1}_{0})_{12}(\omega-(t^{1}_{0})_{11}^{*})}{\square}, \label{4.48}\\
p^{[1]}&=& \frac{p(\omega-(t^{1}_{0})_{11})^{2}+p^{*}(t^{1}_{0})_{12}^{2}-2\eta (t^{1}_{0})_{12}(\omega-(t^{1}_{0})_{11})}{\square}, \label{4.49}\\
{p^{*}}^{[1]}&=&\frac{p^{*}(\omega-(t^{1}_{0})_{11}^{*})^{2}+p(t^{1}_{0})_{12}^{*2}
-2\eta(t^{1}_{0})_{12}^{*}(\omega-(t^{1}_{0})_{11}^{*})}{\square}. \label{4.50}
\end{eqnarray}

\subsection{N-fold DT}

To construct  the $n$-fold DT  of the (2+1)-dimensional SMBE, we introduce $n$ eigenfunctions \cite{C1}
\[\left(\begin{matrix}\Phi_{1,i}\\
 \Phi_{2,i}\end{matrix}\right)=\Phi|_{\lambda=\lambda_i},\, i=1,2,\dots,2n\]
where $\lambda_i$ and $\Phi_i$ satisfy the conditions: $\lambda_{2n-1}=\lambda_{2n}^*$ and   $\Phi_{2,2n}=\Phi_{1,2n-1}^*$ ,  $ \Phi_{2,2n-1}=-\Phi_{1,2n}^*$. 
The $n$-fold DT of the (2+1)-dimensional SMBE can be written as (these formulas are same as for the Hirota-Maxwell-Bloch equation)     \cite{C1}:
\begin{eqnarray}
T_{nt}+T_nU&=&U^{[n]}T_n, \label{4.59}\\
T_{nz}+T_nV&=&V^{[n]}T_n. \label{4.60}
\end{eqnarray}
Hence we obtain 
\begin{eqnarray}
U_{0}^{[n]}&=&U_0+i[\sigma_3,t_{n-1}^{[n]}], \label{4.61}\\
V_{-1}^{[n]}&=&T_n|_{\lambda=-\omega} V_{-1}T_n^{-1}|_{\lambda=-\omega}. \label{4.62}\end{eqnarray}
The matrix  $T_n$ can be expand as \cite{C1}:
\begin{eqnarray}T_n(\lambda;\lambda_1,\lambda_2,\lambda_3,\lambda_4,\dots,\lambda_{2n})=\lambda^n I+t_{n-1}^{[n]}\lambda^{n-1}+\dots+t_1^{[n]}\lambda+t_0^{[n]}=\frac{1}{\Delta_n}\left(\begin{matrix}T_{n11}&T_{n12}\\
T_{n21}&T_{n22}
\end{matrix}\right), \label{4.52}
\end{eqnarray}
where
\begin{eqnarray} \Delta_n&=&
\tiny\left|\begin{matrix}\Phi_{1,1}&\Phi_{2,1}&\lambda_1\Phi_{1,1}&\lambda_1\Phi_{2,1}&\dots&\lambda_1^{n-1}\Phi_{1,1}&\lambda_1^{n-1}\Phi_{2,1}\\
\Phi_{1,2}&\Phi_{2,2}&\lambda_2\Phi_{1,2}&\lambda_2\Phi_{2,2}&\dots&\lambda_2^{n-1}\Phi_{1,2}&\lambda_2^{n-1}\Phi_{2,2}\\
\Phi_{1,3}&\Phi_{2,3}&\lambda_3\Phi_{1,3}&\lambda_3\Phi_{2,3}&\dots&\lambda_3^{n-1}\Phi_{1,3}&\lambda_3^{n-1}\Phi_{2,3}\\
\Phi_{1,4}&\Phi_{2,4}&\lambda_4\Phi_{1,4}&\lambda_4\Phi_{2,4}&\dots&\lambda_4^{n-1}\Phi_{1,4}&\lambda_4^{n-1}\Phi_{2,4}\\
\vdots&\vdots&\vdots&\vdots&\vdots&\vdots&\vdots\\
\Phi_{1,2n-1}&\Phi_{2,2n-1}&\lambda_{2n-1}\Phi_{1,2n-1}&\lambda_{2n-1}\Phi_{2,2n-1}&\dots&\lambda_{2n-1}^{n-1}\Phi_{1,2n-1}&\lambda_{2n-1}^{n-1}\Phi_{2,2n-1}\\
\Phi_{1,2n}&\Phi_{2,2n}&\lambda_{2n}\Phi_{1,2n}&\lambda_{2n}\Phi_{2,2n}&\dots&\lambda_{2n}^{n-1}\Phi_{1,2n}&\lambda_{2n}^{n-1}\Phi_{2,2n}\\
\end{matrix}\right| \label{4.53}
\end{eqnarray}
\begin{eqnarray}T_{n11}=
\tiny\left|\begin{matrix}1&0&\lambda&0&\dots&\lambda^{n-1}&0&\lambda^n\\
\Phi_{1,1}&\Phi_{2,1}&\lambda_1\Phi_{1,1}&\lambda_1\Phi_{2,1}&\dots&\lambda_1^{n-1}\Phi_{1,1}&\lambda_1^{n-1}\Phi_{2,1}&\lambda_1^{n}\Phi_{1,1}\\
\Phi_{1,2}&\Phi_{2,2}&\lambda_2\Phi_{1,2}&\lambda_2\Phi_{2,2}&\dots&\lambda_2^{n-1}\Phi_{1,2}&\lambda_2^{n-1}\Phi_{2,2}&\lambda_2^{n}\Phi_{1,2}\\
\Phi_{1,3}&\Phi_{2,3}&\lambda_3\Phi_{1,3}&\lambda_3\Phi_{2,3}&\dots&\lambda_3^{n-1}\Phi_{1,3}&\lambda_3^{n-1}\Phi_{2,3}&\lambda_3^{n}\Phi_{1,3}\\
\Phi_{1,4}&\Phi_{2,4}&\lambda_4\Phi_{1,4}&\lambda_4\Phi_{2,4}&\dots&\lambda_4^{n-1}\Phi_{1,4}&\lambda_4^{n-1}\Phi_{2,4}&\lambda_4^{n}\Phi_{1,4}\\
\vdots&\vdots&\vdots&\vdots&\vdots&\vdots&\vdots&\vdots\\
\Phi_{1,2n-1}&\Phi_{2,2n-1}&\lambda_{2n-1}\Phi_{1,2n-1}&\lambda_{2n-1}\Phi_{2,2n-1}&\dots&\lambda_{2n-1}^{n-1}\Phi_{1,2n-1}&\lambda_{2n-1}^{n-1}\Phi_{2,2n-1}&\lambda_{2n-1}^{n}\Phi_{1,2n-1}\\
\Phi_{1,2n}&\Phi_{2,2n}&\lambda_{2n}\Phi_{1,2n}&\lambda_{2n}\Phi_{2,2n}&\dots&\lambda_{2n}^{n-1}\Phi_{1,2n}&\lambda_{2n}^{n-1}\Phi_{2,2n}&\lambda_{2n}^{n}\Phi_{1,2n}
\end{matrix}\right| \label{4.54}
\end{eqnarray}
\begin{eqnarray}T_{n12}=
\tiny\left|\begin{matrix}0&1&0&\lambda&\dots&0&\lambda^{n-1}&0\\
\Phi_{1,1}&\Phi_{2,1}&\lambda_1\Phi_{1,1}&\lambda_1\Phi_{2,1}&\dots&\lambda_1^{n-1}\Phi_{1,1}&\lambda_1^{n-1}\Phi_{2,1}&\lambda_1^{n}\Phi_{1,1}\\
\Phi_{1,2}&\Phi_{2,2}&\lambda_2\Phi_{1,2}&\lambda_2\Phi_{2,2}&\dots&\lambda_2^{n-1}\Phi_{1,2}&\lambda_2^{n-1}\Phi_{2,2}&\lambda_2^{n}\Phi_{1,2}\\
\Phi_{1,3}&\Phi_{2,3}&\lambda_3\Phi_{1,3}&\lambda_3\Phi_{2,3}&\dots&\lambda_3^{n-1}\Phi_{1,3}&\lambda_3^{n-1}\Phi_{2,3}&\lambda_3^{n}\Phi_{1,3}\\
\Phi_{1,4}&\Phi_{2,4}&\lambda_4\Phi_{1,4}&\lambda_4\Phi_{2,4}&\dots&\lambda_4^{n-1}\Phi_{1,4}&\lambda_4^{n-1}\Phi_{2,4}&\lambda_4^{n}\Phi_{1,4}\\
\vdots&\vdots&\vdots&\vdots&\vdots&\vdots&\vdots&\vdots\\
\Phi_{1,2n-1}&\Phi_{2,2n-1}&\lambda_{2n-1}\Phi_{1,2n-1}&\lambda_{2n-1}\Phi_{2,2n-1}&\dots&\lambda_{2n-1}^{n-1}\Phi_{1,2n-1}&\lambda_{2n-1}^{n-1}\Phi_{2,2n-1}&\lambda_{2n-1}^{n}\Phi_{1,2n-1}\\
\Phi_{1,2n}&\Phi_{2,2n}&\lambda_{2n}\Phi_{1,2n}&\lambda_{2n}\Phi_{2,2n}&\dots&\lambda_{2n}^{n-1}\Phi_{1,2n}&\lambda_{2n}^{n-1}\Phi_{2,2n}&\lambda_{2n}^{n}\Phi_{1,2n}
\end{matrix}\right| \label{4.55}
\end{eqnarray}

\begin{eqnarray}T_{n21}=
\tiny\left|\begin{matrix}1&0&\lambda&0&\dots&\lambda^{n-1}&0&0\\
\Phi_{1,1}&\Phi_{2,1}&\lambda_1\Phi_{1,1}&\lambda_1\Phi_{2,1}&\dots&\lambda_1^{n-1}\Phi_{1,1}&\lambda_1^{n-1}\Phi_{2,1}&\lambda_1^{n}\Phi_{2,1}\\
\Phi_{1,2}&\Phi_{2,2}&\lambda_2\Phi_{1,2}&\lambda_2\Phi_{2,2}&\dots&\lambda_2^{n-1}\Phi_{1,2}&\lambda_2^{n-1}\Phi_{2,2}&\lambda_2^{n}\Phi_{2,2}\\
\Phi_{1,3}&\Phi_{2,3}&\lambda_3\Phi_{1,3}&\lambda_3\Phi_{2,3}&\dots&\lambda_3^{n-1}\Phi_{1,3}&\lambda_3^{n-1}\Phi_{2,3}&\lambda_3^{n}\Phi_{2,3}\\
\Phi_{1,4}&\Phi_{2,4}&\lambda_4\Phi_{1,4}&\lambda_4\Phi_{2,4}&\dots&\lambda_4^{n-1}\Phi_{1,4}&\lambda_4^{n-1}\Phi_{2,4}&\lambda_4^{n}\Phi_{2,4}\\
\vdots&\vdots&\vdots&\vdots&\vdots&\vdots&\vdots&\vdots\\
\Phi_{1,2n-1}&\Phi_{2,2n-1}&\lambda_{2n-1}\Phi_{1,2n-1}&\lambda_{2n-1}\Phi_{2,2n-1}&\dots&\lambda_{2n-1}^{n-1}\Phi_{1,2n-1}&\lambda_{2n-1}^{n-1}\Phi_{2,2n-1}&\lambda_{2n-1}^{n}\Phi_{2,2n-1}\\
\Phi_{1,2n}&\Phi_{2,2n}&\lambda_{2n}\Phi_{1,2n}&\lambda_{2n}\Phi_{2,2n}&\dots&\lambda_{2n}^{n-1}\Phi_{1,2n}&\lambda_{2n}^{n-1}\Phi_{2,2n}&\lambda_{2n}^{n}\Phi_{2,2n}
\end{matrix}\right| \label{4.56}
\end{eqnarray}
\begin{eqnarray}T_{n22}=
\tiny\left|\begin{matrix}0&1&0&\lambda&\dots&0&\lambda^{n-1}&\lambda^n\\
\Phi_{1,1}&\Phi_{2,1}&\lambda_1\Phi_{1,1}&\lambda_1\Phi_{2,1}&\dots&\lambda_1^{n-1}\Phi_{1,1}&\lambda_1^{n-1}\Phi_{2,1}&\lambda_1^{n}\Phi_{2,1}\\
\Phi_{1,2}&\Phi_{2,2}&\lambda_2\Phi_{1,2}&\lambda_2\Phi_{2,2}&\dots&\lambda_2^{n-1}\Phi_{1,2}&\lambda_2^{n-1}\Phi_{2,2}&\lambda_2^{n}\Phi_{2,2}\\
\Phi_{1,3}&\Phi_{2,3}&\lambda_3\Phi_{1,3}&\lambda_3\Phi_{2,3}&\dots&\lambda_3^{n-1}\Phi_{1,3}&\lambda_3^{n-1}\Phi_{2,3}&\lambda_3^{n}\Phi_{2,3}\\
\Phi_{1,4}&\Phi_{2,4}&\lambda_4\Phi_{1,4}&\lambda_4\Phi_{2,4}&\dots&\lambda_4^{n-1}\Phi_{1,4}&\lambda_4^{n-1}\Phi_{2,4}&\lambda_4^{n}\Phi_{2,4}\\
\vdots&\vdots&\vdots&\vdots&\vdots&\vdots&\vdots&\vdots\\
\Phi_{1,2n-1}&\Phi_{2,2n-1}&\lambda_{2n-1}\Phi_{1,2n-1}&\lambda_{2n-1}\Phi_{2,2n-1}&\dots&\lambda_{2n-1}^{n-1}\Phi_{1,2n-1}&\lambda_{2n-1}^{n-1}\Phi_{2,2n-1}&\lambda_{2n-1}^{n}\Phi_{2,2n-1}\\
\Phi_{1,2n}&\Phi_{2,2n}&\lambda_{2n}\Phi_{1,2n}&\lambda_{2n}\Phi_{2,2n}&\dots&\lambda_{2n}^{n-1}\Phi_{1,2n}&\lambda_{2n}^{n-1}\Phi_{2,2n}&\lambda_{2n}^{n}\Phi_{2,2n}
\end{matrix}\right|. \label{4.57}
\end{eqnarray}

Using  the $n$-fold DT, finally we can give the  determinant form of the $n$-th solution of the (2+1)-dimensional SMBE. It is given by
\begin{eqnarray}
q^{[n]}&=&q+2i(t_{n-1}^{[n]})_{12}, \label{4.63}\\ 
v^{[1]}& = &v-4i(t^{[n]}_{n-1})_{11y}, \label{4.47}\\
p^{[n]}&=&\frac{2\eta T_{n11}T_{n12}
-p^*T_{n12}^{2}
+pT_{n11}^{2}}{T_{n11}T_{n22}-T_{n12}T_{n21}}|_{\lambda=-\omega}, \label{4.64}\\
\eta^{[n]}&=&\frac{\eta(T_{n11}T_{n22}+T_{n12}T_{n21}) -p^*T_{n12}T_{n22}
+pT_{n11}T_{n21}}{T_{n11}T_{n22}-T_{n12}T_{n21}}|_{\lambda=-\omega}. \label{4.65}
\end{eqnarray}
 Here   we considered the  $n$-fold DT  of  the (2+1)-dimensional SMBE.   In the next sections, we apply this DT to construct some exact solutions of this equation.

\section{Solutions}

\subsection{Periodic solutions}

First let us consider a periodic solution of the (2+1)-dimensional SMBE. To do it, as a seed solution,  we  take the following its solution $q=de^{i\rho}, \quad v=m,\quad  p=ifq, \quad \eta=1$. Here $\rho=ax+by+ct$ and $a, b, c, m, f, d$ are some constants. Then  the corresponding eigenfunctions are given by (see i.e.  Refs.\cite{He1}-\cite{Shan})
\begin{eqnarray}
\psi_{1}(\lambda;x,y,t) &=& d e^{\frac{i}{2}\rho+ic(\lambda)}, \label{5.1}\\
\psi_{2}(\lambda;x,y,t)&=& (i(c_{1}\lambda+\frac{b}{2})-\lambda) e^{(-\frac{i}{2}\rho+ic(\lambda)) },\label{5.2}
\end{eqnarray}  
where
$c(\lambda)=c_1(\lambda)t+c_{2}(\lambda)x+c_{3}(\lambda)y$. Now using Eqs.(\ref{5.1}-(\ref{5.2}) we can write  the  periodic solution of the (2+1)-dimensional SMBE corresponding to the one-fold DT. 

\subsection{Soliton solutions}

 To get the one-soliton solution we take the seed solution as  $q=0$, $v=0$, $p=0$, $\eta=1$. Let  $\lambda_{1}=a+bi$. Then the corresponding associated linear system takes the form
 \begin{eqnarray}
\Psi_{1x}&=&-i\lambda \Psi_1,\label{5.6}\\
\Psi_{2x}&=&i\lambda \Psi_2,\label{5.7}\\
\Psi_{1t}&=&2\lambda\Psi_{1y}+\frac{i}{\lambda+\omega}\Psi_1,\label{5.8} \\
\Psi_{2t}&=&2\lambda\Psi_{2y}-\frac{i}{\lambda+\omega}\Psi_2.\label{5.9} 
\end{eqnarray}  
This system admits the following exact solutions
\begin{eqnarray}
\Psi_{1}&=&\Psi_{10}e^{-i\lambda_{1} x+i\mu_{1} y+i(2\lambda_{1}\mu_{1}+\frac{1}{\lambda_{1}+\omega})t},\label{5.10}\\
\Psi_{2}&=&\Psi_{20}e^{i\lambda_{1} x-i\mu_{1} y-i(2\lambda_{1}\mu_{1}+\frac{1}{\lambda_{1}+\omega})t},\label{5.11}
\end{eqnarray}
or
\begin{eqnarray}
\Psi_{1}&=&e^{-i\lambda_{1} x+i\mu_{1} y+i(2\lambda_{1}\mu_{1}+\frac{1}{\lambda_{1}+\omega})t+\delta_1+i\delta_2},\label{5.12}\\
\Psi_{2}&=&e^{i\lambda_{1} x-i\mu_{1} y-i(2\lambda_{1}\mu_{1}+\frac{1}{\lambda_{1}+\omega})t-\delta_1-i\delta_2+i\delta_0},\label{5.13}
\end{eqnarray}
where $\mu_{1}=c+id$, $\delta_i$ and $c,d$ are real constants.  Then  the one-soliton solution of the (2+1)-dimensional SMBE  is given by
\begin{eqnarray}
q^{[1]} &=&   \frac{-4 ai e^{L1}}{e^{N1}+e^{N2}}, \label{5.14}\\
v^{[1]}& = &v-4i(t^{1}_{0})_{11y}, \label{4.47}\\
p^{[1]} &=& \frac{4a e^{L1}[(\lambda_{1}+\omega)e^{N1}-(\lambda_{1}^{*}-\omega)e^{N2}]}{(e^{N1}+e^{N2})^{2}},\label{5.15}\\
\eta^{[1]} &=& \frac{4a^{2} e^{L1+L2}[(\lambda_{1}+\omega)e^{N1}-(\lambda_{1}^{*}-\omega)e^{N2}]}{(e^{N1}+e^{N2})^{2}}
\frac{[(\lambda_{1}+\omega)e^{N2}-(\lambda_{1}^{*}-\omega)e^{N1}]}{(e^{N1}+e^{N2})^{2}},\label{5.16}
\end{eqnarray}
where\\
$L1=2bx-2dy-4(bc+ad)t+\frac{2(a+\omega)}{(a-bi+\omega)(a+bi+\omega)}ti+\delta_{0}i$;\\
$L2=-2bx+2dy-4(bc+ad)t-\frac{2(a+\omega)}{(a-bi+\omega)(a+bi+\omega)}ti+\delta_{0}i$;\\
$N1=-2ax+2cyi+4(ac-bd)ti+\frac{2(a+\omega)}{(a-bi+\omega)(a+bi+\omega)}ti+2(\delta_{1}+\delta_{2}i)$;\\
$N2=2ax-2cyi-(4(ac-bd)ti+\frac{2(a+\omega)}{(a-bi+\omega)(a+bi+\omega)})ti-2(\delta_{1}+\delta_{2}i-\delta_{0}i)$.

Using the above presented  $n$-fold DT, similarly we can construct the $n$-soliton solution of  the (2+1)-dimensional SMBE. 

\section{Conclusion}
 In this paper, we have constructed the DT for the (2+1)-dimensional SMBE. Using the derived DT, some exact solutions like, one-soliton solution and periodic solution are obtained. The determinant  representations of the obtained solutions of the (2+1)-dimensional SMBE are given. Using the above presented results, one can also construct the $n$-solitons, breathers and rogue wave solutions of  the (2+1)-dimensional SMBE.  It is interesting to note that rogue wave solutions of nonlinear equations is currently one of the hottest topics in nonlinear physics and mathematics. The application of the obtained solutions in physics will be one interesting subject. In particular, we hope that the presented solutions may find some usages in experiment or optical fibre communication. Also we note that we will study some important generalizations of the (2+1)-dimensional SMBE in future.

\end{document}